\pdfoutput=1
\documentclass[CEJP,PDF]{cej} 
\usepackage{layout}
\usepackage{amsmath}
\usepackage{textcomp}
\usepackage{hyperref}
\usepackage{graphicx}
\usepackage{subfigure}
\usepackage{color}
\usepackage{xspace}

\newcommand{\twofigs}{0.4\linewidth}

\def\Fig#1{Fig.~\ref{#1}}

\definecolor{kopie}{rgb}{0.2,0.2,0.6}

\graphicspath{
{./}
{../../figures/}
}

\title{Higher-order ratios of baryon number
  cumulants}

\articletype{Review Article} 

\author{Bernd-Jochen
  Schaefer\inst{1}$^,$\inst{2}\email{bernd-jochen.schaefer@uni-graz.at}
  and
  Mathias Wagner\inst{3}\email{mwagner@physik.uni-bielefeld.de}}

\institute{
     \inst{1}  Institut f\"ur Theoretische Physik,
     Justus-Liebig-Universit\"at Gie{\ss}en, Heinrich-Buff-Ring 16, 35392 Gie{\ss}en, Germany
     \inst{2}  Institut f\"ur Physik, 
     Karl-Franzens-Universit\"at Graz, Universit\"atsplatz 5, 8010 Graz, Austria
     \inst{3} Fakult\"at f\"ur Physik,
      Universit\"at Bielefeld, Universit\"atsstra\ss{}e, 33615 Bielefeld, Germany}

    \abstract{The relevance of higher order cumulants of net baryon
      number fluctuations for the analysis of freeze-out and critical
      conditions in heavy-ion collisions at LHC and RHIC is addressed.
      The sign structure of the higher order cumulants in the vicinity
      of the chiral crossover temperature might be a sensitive probe
      and may allow to elucidate their relation to the QCD phase
      transition. We calculate ratios of generalized quark-number
      susceptibilities to high orders in three flavor QCD-like models
      and investigate their sign structure close to the chiral
      crossover line.}


\begin{document}
\maketitle


\section{Introduction}

One of the distinctive feature of the QCD phase diagram is the
possible emergence of a critical endpoint (CEP) where the first-order
chiral phase transition line at large chemical potential
terminates. Many properties of the endpoint such as, e.g., its precise
location are still unknown. However, for experiments that search for
this point the knowledge of its characteristics is inevitable. Since
the expected transition is of second-order at this point possible
signatures are based on the singular behavior of the thermodynamic
functions in its vicinity.  Recently, it was pointed out that
event-by-event fluctuations of particle multiplicities and their
nonmonotonic behavior might serve as a probe to locate a possible CEP
in the phase diagram.  By scanning the center of mass energy and thus
the baryochemical potential an increase and then a decrease in the
number fluctuations of, e.g., pions and protons should be seen as one
crosses the critical point. Unfortunately, in a realistic heavy-ion
collision the expected signals are washed out due to the critical
slowing down phenomenon and finite volume effects. Furthermore,
correlations can only build up for a finite time in the colliding
system and consequently the correlations length is cut off. Some
conservative estimates yield an increase of the correlation length
only by a factor 3 as the system crosses the critical
point. Therefore, more sensible quantities are needed for the analysis
of the freeze-out and critical conditions in heavy-ion collisions. For
this reason higher-order cumulants or ratios of higher-order
generalized susceptibilities have been suggested as suitable
quantities because they depend on higher powers of the correlation
length \cite{Stephanov:1999zu}.

\section{A three-flavor model analysis}

\begin{figure}
\centering
  \includegraphics[width=\twofigs]{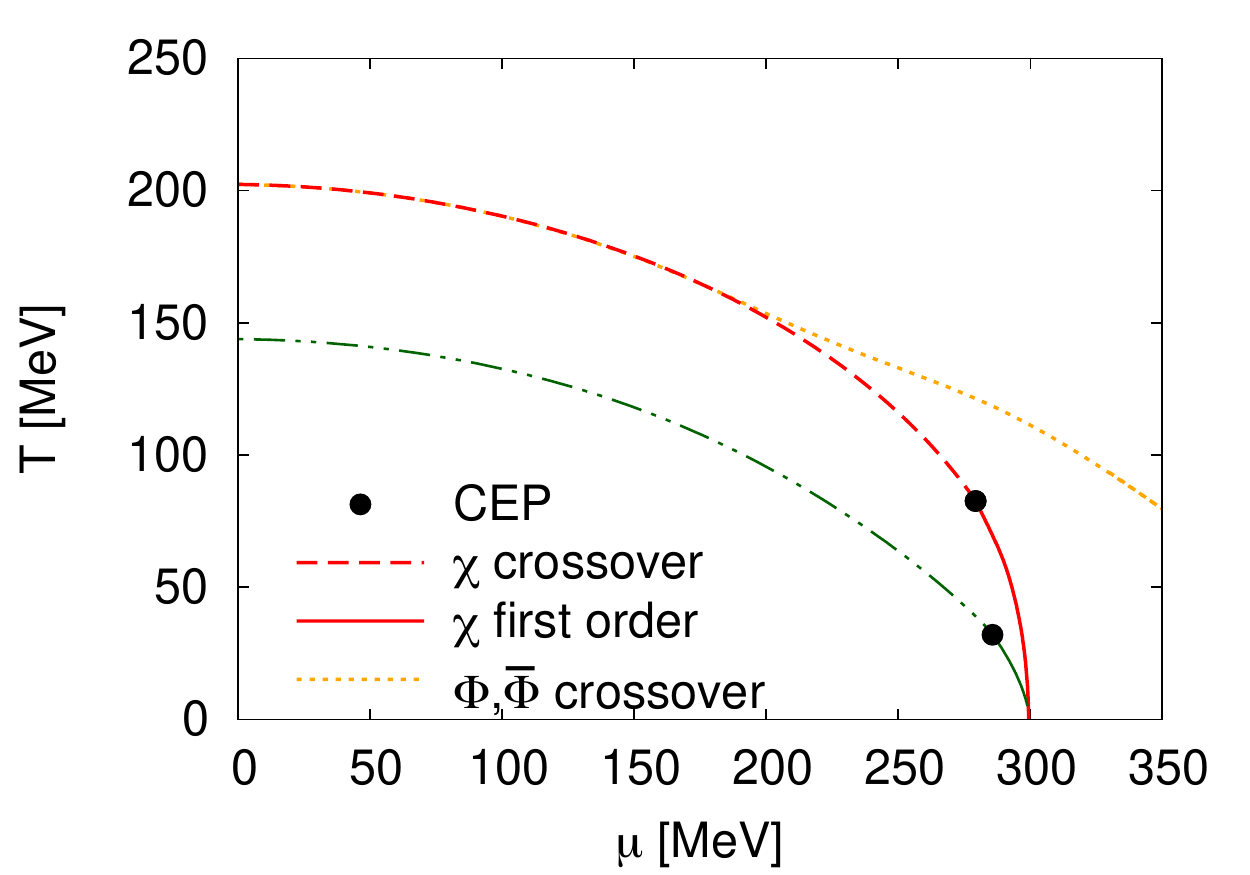}
\caption{\label{fig:pqmpd}Phase diagrams for three flavor of the
  renormalized PQM model with a running $T_0$ parameter in the
  logarithmic Polyakov-loop potential and of the quark-meson
  model. Dashed lines denote the chiral crossover, solid lines the
  first-order chiral transition and dotted lines the peak in the
  temperature-derivative of the Polyakov-loops. }
\end{figure}

Fluctuations of conserved charges are quantified by cumulants in
statistics and are related to generalized susceptibilities
\cite{Koch:2008ia}. They are defined as derivatives of the logarithm
of the partition function with respect to the appropriate chemical
potentials. For three different chemical potentials we have
accordingly
\begin{equation}\label{eq:defchi}
\chi_{n_i,n_j,n_k} \equiv \frac{ 1}{VT} \frac{ \partial^{n_i}}{\partial (\mu_{i}/T)^{n_i}} 
\frac{ \partial^{n_j}}{\partial (\mu_{j}/T)^{n_j}} 
\frac{ \partial^{n_k}}{\partial (\mu_{k}/T)^{n_k}} \log Z \ ,
\end{equation}
where $n_i,\ldots$ denotes the order of the derivatives and the
indices $(i,j,k) = (u,d,s)$ the quark flavor. The generalized
susceptibilities evaluated at vanishing $\mu_f$ are the Taylor
expansion coefficients of the pressure series in powers of $\mu_f/T$.
The partition function is evaluated in a renormalized three-flavor
Polyakov-quark-meson (PQM) model with an axial $U(1)_A$ symmetry
breaking term and a logarithmic Polyakov-loop potential in mean-field
approximation where the ultraviolet divergent fermion vacuum
contribution has been taken into account \cite{Schaefer:2011ex}. Thus,
fermion fluctuations are taken into account whereas meson and
Polyakov-loop fluctuations are still ignored. The resulting phase
diagram in comparison with the one obtained in the three-flavor
quark-meson model is shown in \Fig{fig:pqmpd}

\subsection{Ratios of baryon number moments}

In the following we focus on one uniform quark chemical
potential and denote the $n$th to $m$th order moment ratio of the
quark number fluctuations $\chi_n$ as
\begin{equation}
  R_{n,m}\equiv \chi_n (T,\mu)/\chi_m (T,\mu)\ .
\end{equation}
The evaluation of the ratios has been automated by using algorithmic
differentiation techniques \cite{Wagner:2009pm}.  The kurtosis
$R_{4,2}$ basically measures the quark content of particles carrying
baryon number and has been calculated in the vicinity of the crossover
region of the phase diagram.  In the hadron resonance gas (HRG) model
the ratio is temperature independent and all moments stay positive
since the HRG model has no singularities.  Thus, any deviation from
the HRG model result might be an indicator for a real critical
phenomenon even if the lower-order moments and the thermodynamics
including particle yields are well described by the HRG
model~\cite{Andronic:2009qf}. In our model calculation the kurtosis
becomes negative and consequently a negative region around the
crossover line emerge in the phase diagram.  Note that these
three-flavor results differ from a corresponding two-flavor PQM
mean-field calculation with and without the fermion vacuum term in the
grand potential. Without the vacuum term in the two-flavor case the
kurtosis $R_{4,2}$ develops a peak at zero chemical potential near the
crossover temperature which is a remnant of the first-order transition
in the chiral limit, see \cite{Nakano:2009ps} for more details. The
inclusion of the fermion vacuum term changes the transition to
second-order in the chiral limit which is consistent with universality
arguments \cite{Skokov:2010sf}. However, for three massless flavor a
first-order transition always emerges in the chiral limit with or
without the fermion vacuum term. The peak of the kurtosis for physical
pion masses is already less pronounced if the vacuum term is neglected
and finally vanishes in the three-flavor renormalized model. Thus, the
suppression of the kurtosis peak at $\mu=0$ must be caused by the
strange quarks and not by fluctuations \cite{Schaefer:2011ex,
  Skokov:2010sf}.

\subsection{Sign structure of higher-order ratios close to the transition}

\begin{figure}
  \centering 
  \subfigure[$\ $Polyakov-quark-meson model]{\label{fig:negreg}
    \includegraphics[width=\twofigs]{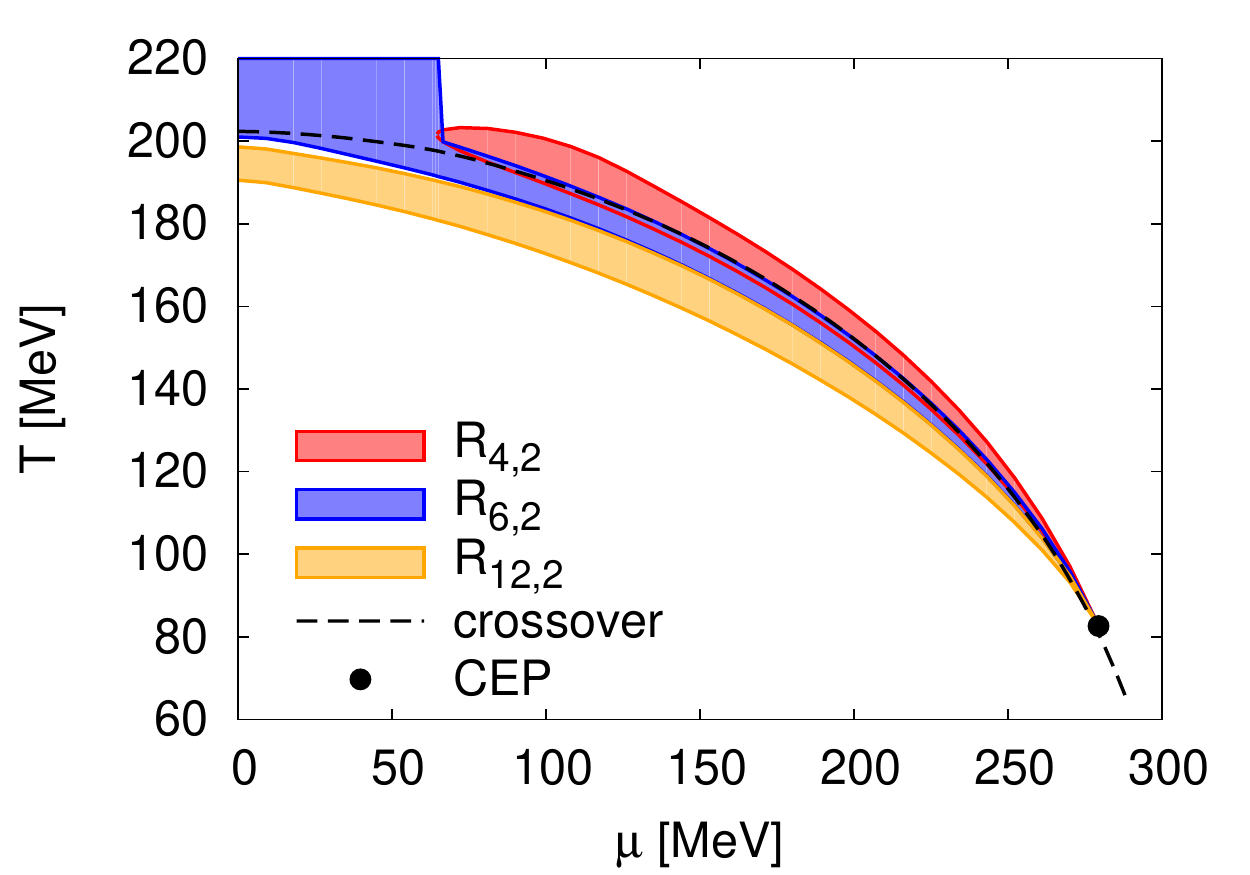}}
  \subfigure[$\ $Quark-meson model]{
    \includegraphics[width=\twofigs]{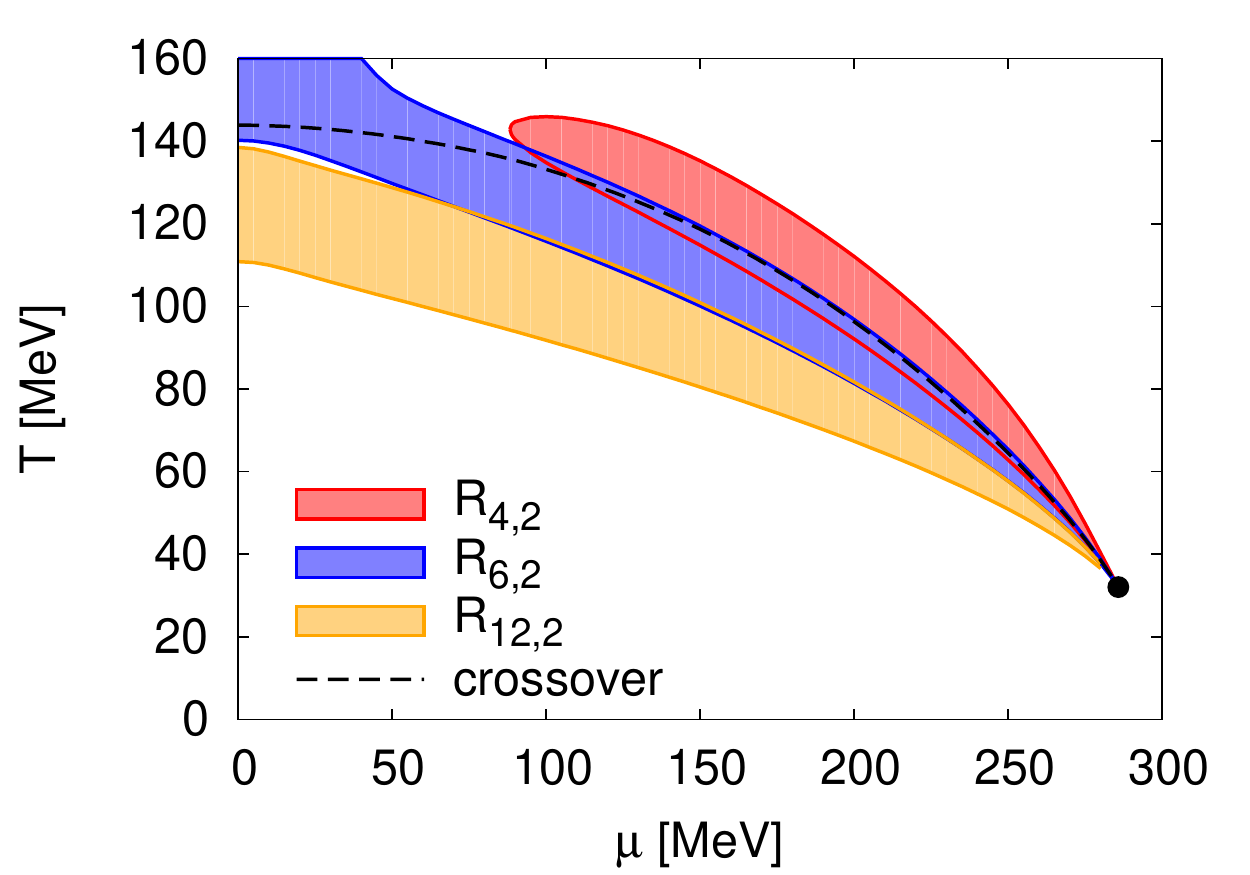}}
  \caption{\label{fig:extremapd}Three negative areas of $R_{n,2}$
    ratios close to the chiral crossover line in both models.}
\end{figure}

Higher-order ratios behave in a similar way. They oscillate within a
narrow temperature interval for temperatures close to the chiral
transition. Generically, the structures of all moments at $\mu=0$ are
related to each other and the behavior including the amplitudes of
$\chi_n$ can be deduced by the temperature derivative of the preceding
$\chi_{n-2}$~\cite{Schaefer:2009st}. In contrast to the HRG model all
higher-order ratios become negative in the vicinity of the crossover
line for nonvanishing chemical potential as shown in
\Fig{fig:extremapd}. In this figure the negative regions of three even
ratios $R_{n,2}$ in the renormalized three-flavor PQM model along the
chiral crossover line are compared to a corresponding three-flavor
renormalized QM model without the Polyakov-loop.  All negative regions
are closely correlated with the crossover curve and converge exactly
at the CEP. For $n>4$ they are shifted slightly in the hadronic phase.

The Polyakov loop sharpens these negative regions around the chiral
transition.  In the renormalized models the CEP is pushed towards
higher chemical potentials by the inclusion of the vacuum terms but
the curvature of the crossover line seems not to be changed. However,
the crossover is washed out by fluctuations which yields larger
negative regions.
In summary the behavior of the negative regions can surely be
attributed to critical dynamics~\cite{Skokov:2011rq}. The findings
underline once more the importance of fluctuations: all regions
calculated in the renormalized models are shifted more in the hadronic
phase.

\begin{figure*}
  \centering
    \includegraphics[width=\twofigs]{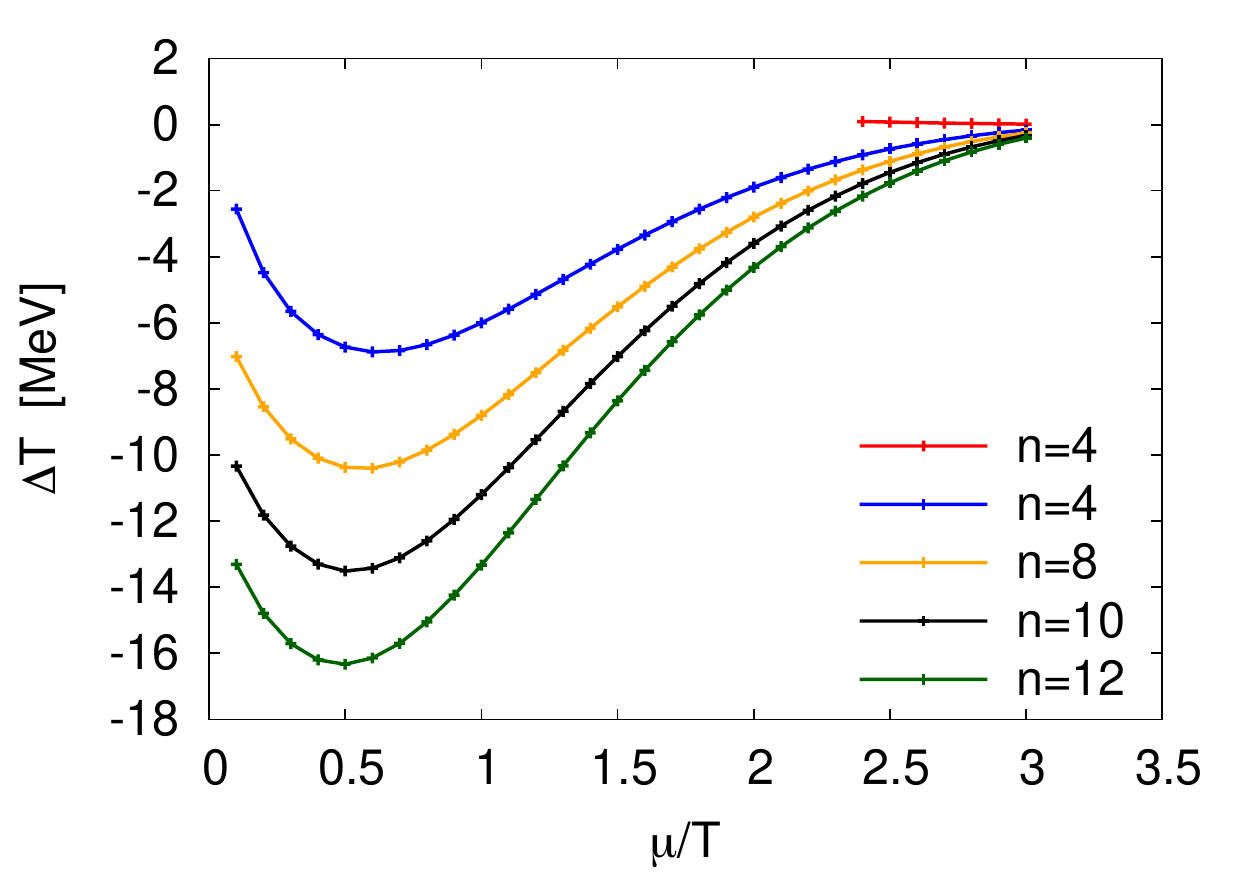}
\includegraphics[width=\twofigs]{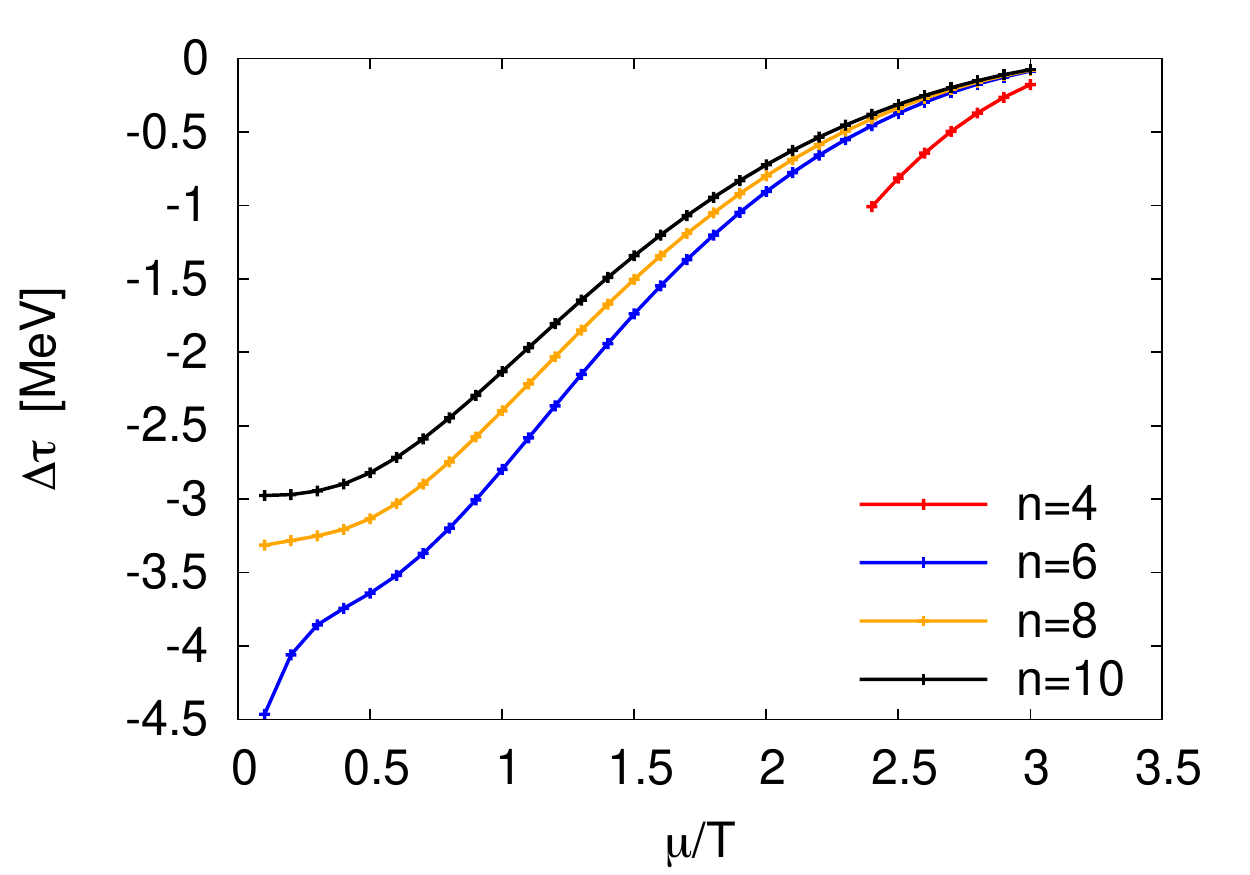} 
\caption{\label{fig:oddroots_pb}Distance $\Delta T = T_n - T_\chi$
  (left) and
  the relative distance $\Delta \tau= T_{n+2} - T_n$ (right) of the first zero
  for various ratios $R_{n,2}$ for even integers $n$ as a function of
  $\mu/T$ in the renormalized PQM model. $T_\chi$ denotes the chiral
  critical temperature.}
\end{figure*}

The knowledge of how the negative regions evolve towards the endpoint
might be used to construct new criteria to improve the critical
temperature estimate from a Taylor expansion around $\mu=0$. For this
reason it is instructive to define the distance $\Delta T = T_n -
T_\chi$ of the first zero in temperature direction of the ratio
$R_{n,2}$ to the crossover temperature $T_\chi$. In the left panel of
\Fig{fig:oddroots_pb} the distance $\Delta T$ of several even ratios
$R_{n,2}$ is shown as a function of $\mu/T$. Only the ratio $R_{4,2}$
remains positive away from the CEP. For all higher-order ratios the
first zero in $R_{n,2}$ is pushed into the hadronic phase and all
$\Delta T$ are negative already for $\mu/T \sim 0$. They obey a
minimum whose precise location and depth is model dependent. With the
Polyakov loop the transition is sharper and the minima are not as
deep.  Remarkable is the almost linear behavior of $\Delta T$ for
intermediate $\mu/T$ ratios in the PQM and QM models for all ratios
$R_{n,2}$. The linear extrapolation of $\Delta T$ from intermediate
$\mu/T$ to larger values where $\Delta T$ vanishes might serve as an
estimator for the proximity of the thermal freeze-out to the crossover
line and the existence of a possible endpoint can be {\em ruled out}
for smaller values of $\mu/T$.  This estimate could be strengthened by
considering only the difference of the subsequent roots in the odd
ratios which is independent of the knowledge of the chiral crossover
temperature.  The relative temperature distance $\Delta \tau = T_{n+2}
- T_n$ for several even ratios $R_{n,2}$ is shown as a function of
$\mu/T$ in the right panel of \Fig{fig:oddroots_pb}. The curves
exhibit a similar behavior as in the previous figure except the
ordering of the curves with respect to $n$ is opposite. With
increasing $n$ the distance between subsequent ratios decreases,
signaling a possible convergence of the negative regions in the phase
diagram. An estimation of the lower bound of the CEP with a linear fit
is in between 15\% of the actual CEP model values.

\section{Conclusion}

Higher-order cumulants or generalized susceptibilities are more
sensitive on the diverging correlation length and are promising quantities for the experimental search of an endpoint in
the QCD phase diagram.
In our model analysis the higher-order moments differ significantly
from the HRG model results along the freeze-out curve due to the
existence of an endpoint. A region with negative values of the ratios
of momenta emerges close to the chiral transition line.  They are
shifted towards the hadronic phase and converge at the endpoint.  In
order to quantify this general trend we have introduced the distance
of the first zero in temperature direction of various ratios to the
crossover temperature and the relative distance between subsequent
roots which is independent of the insecure chiral crossover
temperature. By using linear extrapolations as estimators we could
rule out the existence of an endpoint for smaller $\mu/T$ ratios.

\end{document}